%% file: ms.tex
\documentclass[useAMS,usenatbib]{mn2e}
\usepackage{epsfig}
\input{macro1}

\title[Clustering of barred galaxies]
{The clustering of barred galaxies in the local Universe}

\author[Li et al.]
{
Cheng Li$^{1,2}$\thanks{E-mail:leech@mpa-garching.mpg.de},
Dimitri A. Gadotti$^{1}$, Shude Mao$^{3}$, Guinevere Kauffmann$^{1}$ \\
$^{1}$Max-Planck-Institute for Astrophysics,
Karl-Schwarzschild-Str. 1, D-85741 Garching, Germany \\
$^{2}$MPA/SHAO Joint Center for Astrophysical Cosmology at Shanghai
Astronomical Observatory, Nandan Road 80, Shanghai 200030, China\\
$^{3}$Jodrell Bank Centre for Astrophysics, Alan Turing Building, 
University of Manchester, Manchester M13 9PL, UK
}

\begin{document}

\date{Accepted ........ Received ........; in original form ........}

\pagerange{\pageref{firstpage}--\pageref{lastpage}} \pubyear{2007}

\maketitle

\label{firstpage}

\begin {abstract}
We study the clustering properties  of barred galaxies using data from
the   Sloan  Digital   Sky  Survey   (SDSS).   We   compute  projected
redshift-space two-point cross-correlation  functions $w_p(r_p)$ for a
sample of  nearly 1000 galaxies  for which we have  performed detailed
structural   decompositions    using   the   methods    described   in
\citet{Gadotti-09}.  The  sample  includes  286 barred  galaxies.  The
clustering of barred and unbarred  galaxies of similar stellar mass is
indistinguishable  over  all   the  scales  probed  ($\sim$20  kpc--30
Mpc). This result also holds even  if the sample is restricted to bars
with bluer  $g-i$ colours  (and hence younger  ages). Our  result also
does not change if we split our sample of barred galaxies according to
bar-to-total  luminosity   ratio,  bar  boxyness,   effective  surface
brightness, length, or the shape of the surface density profile within
the  bar. There is a hint that red, elliptical bars are  more strongly
clustered  than red and less  elliptical  bars, on scales $\ga 1$ Mpc,
although the statistical significance is not high.  We
conclude that there is no significant evidence that bars are a product
of  mergers or  interactions.  We tentatively  interpret the  stronger
clustering  of the  more elliptical  bars  as evidence  that they  are
located in older galaxies, which reside in more massive haloes.

\end {abstract}

\begin{keywords}
galaxies: clustering - galaxies: distances and redshifts - large-scale
structure of Universe - cosmology: theory - dark matter
\end{keywords}

\section {Introduction}
\label{sec:intro}

It has  long been known  that stellar bars  are very common  in spiral
galaxies.  In the  local Universe, about $30\%$ of  disc galaxies have
strong bars; this  fraction increases to $\sim 60\%$  if weak bars are
included (e.g.  \citealt{dev-63, Knapen-00, Marinova-07}).  The bar is
believed to play an important role in triggering the secular evolution
of      the      galaxy      (see     e.g.      \citealt{Sellwood-93};
\citealt{Kormendy-Kennicutt-04}  for  a  review) and  regulating  star
formation (e.g. \citealt{WangJL-06}).

The  redshift  evolution  of   the  bar  fraction  is  still  somewhat
controversial.  Several  early studies of the bar  fraction at $z>0.5$
in the  Hubble Deep Fields (HDFs)  found a dramatic  paucity of barred
galaxies \citep{Abraham-96, Abraham-99, vandenBergh-96}, although this
result    has   been    disputed    by   others    (\citealt{Sheth-03,
  Elmegreen-Elmegreen-Hirst-04, Jogee-04,  Zheng-05}).  An analysis by
\citet{Sheth-08}, which used a  much larger sample, concludes that the
bar  fraction indeed decreases  with redshift,  with the  effect being
stronger for low-mass galaxies.

Two mechanisms  are thought  to contribute to  the formation  of bars.
Bars  can form  through the  $m=2$  mode global  instability in  cold,
rotationally supported  discs. This has been  demonstrated by numerous
numerical                                                   simulations
(e.g. \citealt{Hohl-1971,Ostriker-1973,Sellwood-1981,Athanassoula-1986}).
The  other  formation  mechanism  is through  tidal  perturbations  by
neighbouring  galaxies.    This  has  also   been  demonstrated  using
numerical                                                   simulations
(\citealt{Byrd-1986,Noguchi-1987,Noguchi-1988,Noguchi-1996,Gerin-1990,
  Miwa-1998}).   The effects  induced  by merging  galaxies are  quite
complex.   Simulations of the  secular evolution  of disc  galaxies by
\citet{Debattista-06}  showed  that  interactions  can  speed  up  bar
formation in  direct encounters, but have little  effect in retrograde
ones   \citep{Toomre-72,   Noguchi-87,   Gerin-Combes-Athanassoula-90,
  Steinmetz-Navarro-02,  Aguerri-Gonzalez-Garcia-09}.  Furthermore, it
is possible for mergers to destroy or severely weaken the bar, without
destroying the  disk \citep{Berentzen-2003}.  It is  thus not entirely
clear  what  the  combined  effect  of interactions  and  mergers  is.
Observationally,  examples   of  tidally  triggered   bars  are  known
\citep[e.g.][]{Debattista-02,      Peirani-09}.       In     addition,
\citet{Elmegreen-1990} found an excess  of binary companions in barred
galaxies.

There    is   little   doubt    that   both    internally-driven   and
externally-driven  mechanisms can  trigger the  formation of  bars. We
would like to determine empirically the relative importance of the two
mechanisms,  and also  investigate whether  we can  differentiate bars
formed  through these  two mechanisms.   \cite{Miwa-1998}  studied the
difference between tidally induced  bars and those induced by internal
processes,  and  concluded  that  the  tidal  effects  produce  slowly
rotating bars while those  arising from internal processes rotate much
faster.   Clearly  these theoretical  predictions  need  to be  tested
observationally.

In  this paper,  we  select  carefully a  sample  of nearby ($z<0.07$)
face-on    galaxies    from    the    Sloan   Digital    Sky    Survey
\citep[SDSS;][]{York-00},  determine their structural  properties with
sophisticated imaging fitting, and perform cross-correlation of barred
galaxies  with other  galaxies in  order to  assess the  importance of
tidal interactions for the bar formation.  The outline of the paper is
as follows.  In \S\ref{sec:data},  we describe how our barred galaxies
are   selected   and   how   their  properties   are   measured.    In
\S\ref{sec:methods}  we present  the method  of  our cross-correlation
analysis.  The main results  are presented in \S\ref{sec:results}.  In
\S\ref{sec:summary} we briefly summarise our results and discuss their
implications. Throughout this paper, we use a cosmology with a density
parameter   of  $\Omega_{\rm   m}=0.3$,   and  cosmological   constant
$\Omega_\Lambda=0.7$ and  the Hubble  constant is written  as $H_0=100
h~{\rm km~s^{-1}~Mpc^{-1}}$.

\section{Data}\label{sec:data}

The sample  used for  the classification of  galaxies into  barred and
unbarred  subclasses was drawn  from a  volume-limited sample  in SDSS
data   release    2   \citep[DR2;][]{Abazajian-04},   with   $0.02\leq
z\leq0.07$.   To  ease  the  identification  of bars,  the  sample  is
restricted to galaxies  very close to a face-on  projection, i.e. with
an axial ratio $b/a\geq0.9$, where  $a$ and $b$ are, respectively, the
semi-major and semi-minor axes of the galaxy. We excluded all galaxies
with  stellar  masses   below  $10^{10}~{\rm  M}_\odot$,  since  dwarf
galaxies  are not  object of  our study.   Galaxy stellar  masses were
obtained  from  \citet{KauHecWhi03}.   These  criteria resulted  in  a
sample of  3375 objects.   The images of  each of these  galaxies were
individually inspected  to remove from  the sample galaxies  which are
either  not truly  face-on,  morphologically disturbed,  too faint  or
irregular,  too close  to the  border  of the  CCD frame,  as well  as
duplicate  entries and  images where  the  presence of  a bright  star
hinders  the decomposition.  We also  rejected galaxies  with apparent
size smaller  than $\sim$ 20  pixels across ($\sim$ 8  arcsec), deemed
too small  for parametric  image decomposition.  The  resulting sample
contains 930  galaxies.  We have verified  that this sample  is a fair
representation  of the  galaxy population  in this  mass range  in the
local Universe \citep[see][]{Gadotti-09}.

To verify whether  a galaxy is barred, we  inspected the galaxy image,
isophotal contours and a  pixel-by-pixel radial intensity profile.  We
looked for  typical bar signatures, i.e.  an  elongated structure with
constant position  angle and  a flat ledge  in the  profile \citep[see
  e.g.][]{GadAthCar07}.  It should be  noted that  due to  the limited
spatial resolution of  SDSS images, we miss most  bars with semi-major
axis shorter  than $\approx2-3$  kpc, which are  found mainly  in very
late-type spirals  (later than Sc --  \citealt{ElmElm85}). These faint
bars lie  typically within  $2-4$ seeing elements  and do  not produce
clear  signatures.  The  results presented  here thus  pertain  to the
typical, bonafide bars seen in early-type spirals and lenticulars. For
each  bar in  our sample,  we have  measured a  variety  of properties
through parametric  image fitting.  These  include absolute magnitudes
in $g$, $r$ and $i$ bands, optical colours defined by the three bands,
effective   surface  brightness,   effective  radius,   Sersic  index,
ellipticity,  semi-major axis,  boxyness, and  bar-to-total luminosity
ratio.  Details can be found in \citet{Gadotti-09}.

In this paper we classify  all the galaxies into three types according
to  their  bulge-to-total  luminosity  ratio  $B/T$  and  bar-to-total
luminosity  ratio $b/T$.  A  galaxy is  {\em  elliptical} if  $B/T=1$;
otherwise it is a {\em spiral}  galaxy. In the latter case, the galaxy
is  further classified  as  either  {\em barred}  if  $b/T>0$ or  {\em
  unbarred} if $b/T=0$. This  results in 255 ellipticals, 389 unbarred
and  286  barred  spirals. We  will  use  the  last two  samples  when
comparing barred and un-barred galaxies, and will use the barred sample
on its own for investigating whether bar properties are sensitive
to the environments of their host galaxies.         

\section{Methods}\label{sec:methods}

We probe the clustering properties of our galaxies using the projected
redshift-space    two-point   cross-correlation    function   (2PCCF),
$w_p(r_p)$. We  calculate the cross-correlation between  the sample of
galaxies for which we  have detailed structural information from image
decompositions and  a reference sample  of galaxies selected  from the
Main    spectroscopic   sample    of   the    fourth    data   release
\citep[DR4;][]{Adelman-McCarthy-06} of  the SDSS.  A  random sample is
constructed so as to have  the same selection effects as the reference
sample. The reference and random samples are cross-correlated with the
same set  of galaxies in our  sample, and $w_p(r_p)$ as  a function of
the projected separation $r_p$ is defined by the ratio of the two pair
counts  minus one.  Details  about our  methodology for  computing the
correlation functions  and for  constructing the reference  and random
samples can be found  in \citet{Li-06a}.  For consistency, we restrict
our reference galaxies  to the same redshift range  as our sample with
structural information, that is, $0.02\leq z\leq0.07$.

The amplitude  of 2PCCF  on scales  larger than a  few Mpc  provides a
direct measure  of the mass  of the dark  matter haloes that  host the
galaxies   through  the   halo  mass-bias   relation.   As   shown  in
\citet{Li-08a,Li-08b},  the amplitude of  the correlation  function on
scales $\la 100$  kpc can serve as a probe  of physical processes such
as mergers and interactions.   On intermediate scales, the correlation
probes  the  so-called  ``1-halo''  term  where the  pair  counts  are
dominated by galaxy pairs in the same halo.

\section{Results}\label{sec:results}

\subsection{Comparison between barred and unbarred galaxies}

\begin{figure}
\centerline{\psfig{figure=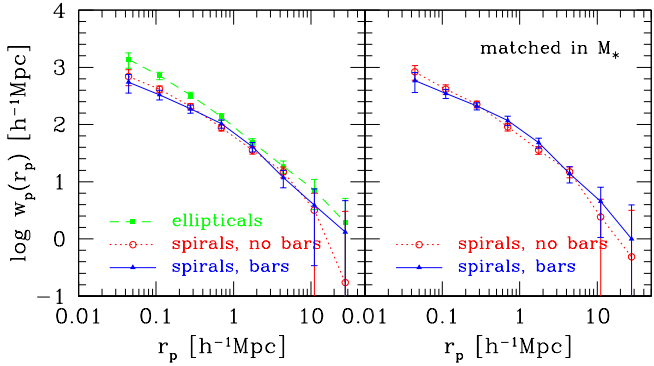,clip=true,width=0.5\textwidth}}
\caption{Projected redshift-space two-point cross-correlation function
  $w_p(r_p)$ for  elliptical galaxies (green squares  connected by the
  dashed line), the unbarred spiral galaxies (red circles connected by
  the dotted line), and  the barred galaxies (blue triangles connected
  by the solid line).  The  right-hand panel shows the results for the
  barred and unbarred spiral samples only, but the samples are closely
  matched in stellar mass.}
\label{fig:wrp}
\end{figure}

\begin{figure}
\centerline{\psfig{figure=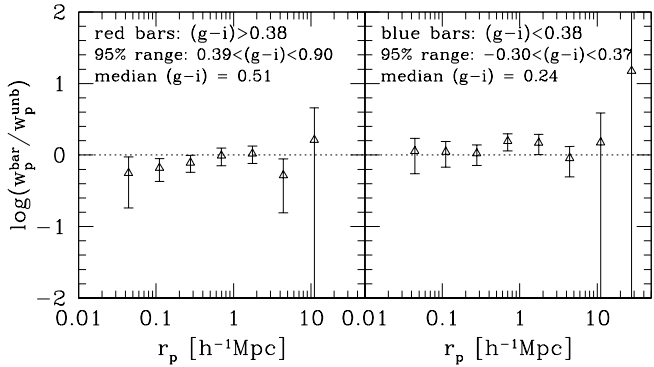,clip=true,width=0.5\textwidth}}
\caption{Ratio  of $w_p(r_p)$  for red  (left panel)  and  blue (right
  panel) bars  relative to that  for unbarred spirals.  In  each panel
  the barred and unbarred samples are closely matched in stellar mass.
  The median and the 95\% range  of the $(g-i)$ colour of the bars are
  indicated.}
\label{fig:wrp_barcolor}
\end{figure}

In  the left-hand  panel of  Fig.~\ref{fig:wrp}, we  compare projected
redshift-space 2PCCF  $w_p(r_p)$ for  our samples of  ellipticals, and
spirals  with and  without bars.   Results for  the three  samples are
plotted as squares, triangles  and circles, respectively, connected by
the dashed, solid and dotted  lines.  This panel shows that elliptical
galaxies  are more  strongly clustered  than  the other  two types  of
galaxies, which is not surprising.   This panel also shows that barred
and unbarred  spirals cluster in  the same way.  The  right-hand panel
shows the  results for the two  spiral samples again, but now the samples
are closely matched  in stellar mass $M_\ast$.  Because clustering  is known to
depend  on $M_\ast$  it  is important  to  control for  this effect  when
investigating the  dependence on  other galaxy properties.   There are
still no significant differences between the two samples.

We tentatively conclude  that the presence of a bar  is not related to
galaxy environment. In order to test whether the same conclusion holds
for  bars that have  formed more  recently, we  thus split  the barred
spirals  into two  equal-size subsamples  according to the $g-i$
colour    of     the    bar. \citet[][and references therein]{Gadotti-deSouza-06} 
suggest that bar colour can be used as a proxy for its dynamical age
\citep[but see][]{Perez-09}.  The    results     are    shown    in
Fig.~\ref{fig:wrp_barcolor}, where we  compare $w_p(r_p)$ for galaxies
with red  (left panel) or blue  (right panel) bars,  together with the
result  for unbarred  spirals.  In  each panel,  the  two samples  are
matched in stellar mass. We conclude that the colour of the bar has no
effect on our main result.

\begin{figure}
\centerline{\psfig{figure=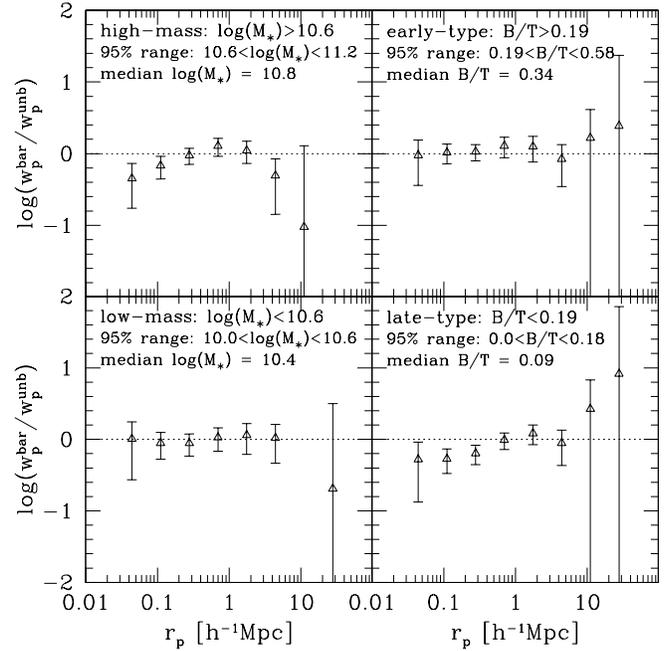,clip=true,width=0.5\textwidth}}
\caption{Ratio  of $w_p(r_p)$ for  all bars  with respect  to unbarred
  spirals.   The  bars  are  divided into  two  equal-size  subsamples
  according  to   either  the  stellar  mass   or  the  bulge-to-total
  luminosity ratio of  their host galaxies, and the  results are shown
  in  the  upper-left  panel  for  the  high-mass  subsample,  in  the
  lower-left for  the low-mass subsample,  in the upper-right  for the
  early-type  subsample,  and in  the  lower-right  for the  late-type
  subsample. In each panel the barred and unbarred samples are closely
  matched in stellar  mass.  The median and the  95\% range of stellar
  mass or bulge-to-total luminosity ratio are indicated.}
\label{fig:wrp_galmass}
\end{figure}

We would  also like to test  whether our result is  independent of the
morphological type of  the galaxy.  For instance, \citet{Giuricin-93b}
found that the spirals in  high density environments tend to be barred
only  if they  are early-type.   We have  split the  barred  sample by
$M_\ast$   and   by   $B/T$,    and   the   results   are   shown   in
Fig.~\ref{fig:wrp_galmass}.  We have also  checked the results for our
subsamples of red and blue bars.   In any case we see no dependence in
clustering.   We thus conclude  that environment  has no  influence on
whether or not bars are present in galaxies.

\subsection{Dependence on bar properties}

\begin{figure*}
\centerline{\psfig{figure=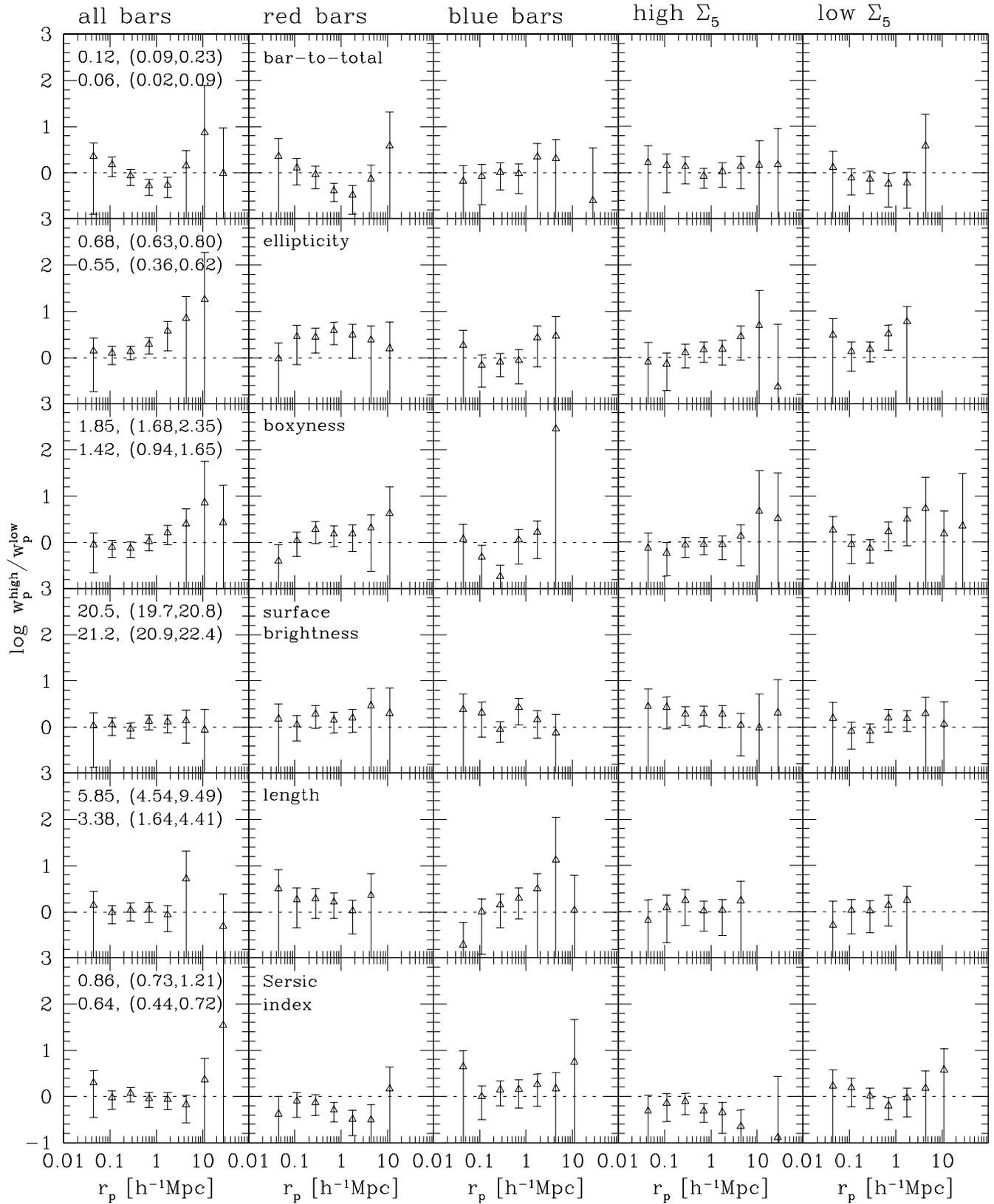,clip=true,width=\textwidth}}
\caption{We test whether the projected redshift-space 2PCCF $w_p(r_p)$
  depends on bar properties.  Panels  from left to right correspond to
  results for  the whole  bar sample, for  subsamples of red  and blue
  bars, and  for subsamples of  bars located in high-  and low-density
  regions. Panels from top to  bottom test the dependence of the 2PCCF
  on the  bar property as indicated. Each panel
  shows  the ratio  of $w_p(r_p)$  between two  subsamples,  which are
  matched  in  stellar  mass,  with  higher and  lower  value  of  the
  property.   The  median and  95\%  range  of  the bar  property  are
  indicated  in  the  left-hand   panels  for  the  two  corresponding
  subsamples.}
\label{fig:wrp_bars}
\end{figure*}

In  this  section  we focus  on  the  barred  galaxies and  study  the
dependence of  clustering on  the properties of  the bars.  As  in the
previous section,  we split  the galaxies into  two subsamples  at the
median value of  the given property.  We also trim  the two samples so
that they each have the same distribution in stellar mass.

The  results are shown  in Fig.~\ref{fig:wrp_bars},  where we  plot in
each  panel the  ratio of  the  measurement of  $w_p(r_p)$ for  barred
galaxies with higher and lower  values of the given bar property.  The
left-hand panels  show results for all  the bars in  our sample, while
the  following  two  panels  are  for  red and  blue  bars.   The  bar
properties  we   consider  are  (from  top   to  bottom)  bar-to-total
luminosity ratio, ellipticity, boxyness, effective surface brightness,
length of the semi-major axis, and Sersic index (see \S~\ref{sec:data}
for detailed description on these quantities).

From Figure~\ref{fig:wrp_bars}  we see some  indications of dependence
of  bar   properties  on  clustering,  however,  they   are  not  very
significant  due to the  large errors  in the  $w_p(r_p)$ measurements
resulting from small sample sizes.  These include the bar ellipticity,
bar-to-total luminosity ratio, and Sersic index for red bars.

Interestingly  there  is  a  hint  of bar  ellipticity  dependence  on
clustering. For  the sample as  a whole, the difference  in clustering
occurs   on  scales   greater   than  $\sim   1  h^{-1}\,{\rm   Mpc}$.
High-ellipticity bars are more strongly clustered than low-ellipticity
bars on  large scales.  When  the bars are  divided into red  and blue
subsamples, the  (weak) effect persists  for the red bars  but largely
vanishes for the blue bars.

One possible  explanation for these  trends (if confirmed  with larger
samples) is  that more  elliptical bars occur  in older,  more evolved
galaxies.  The age of a galaxy of fixed stellar mass will be larger in
more massive haloes than less massive ones, so one would expect to see
clustering differences of the kind that we observe.  The dependence of
bar ellipticity on  age is predicted both from  N-body simulations and
analytical work  \citep[see for example][]{Athanassoula-Misiriotis-02,
  Athanassoula-03}. It would be interesting to examine the ellipticity
dependence on clustering with larger samples in the future.

\subsection{Dependence on local environment}

\begin{figure}
\centerline{\psfig{figure=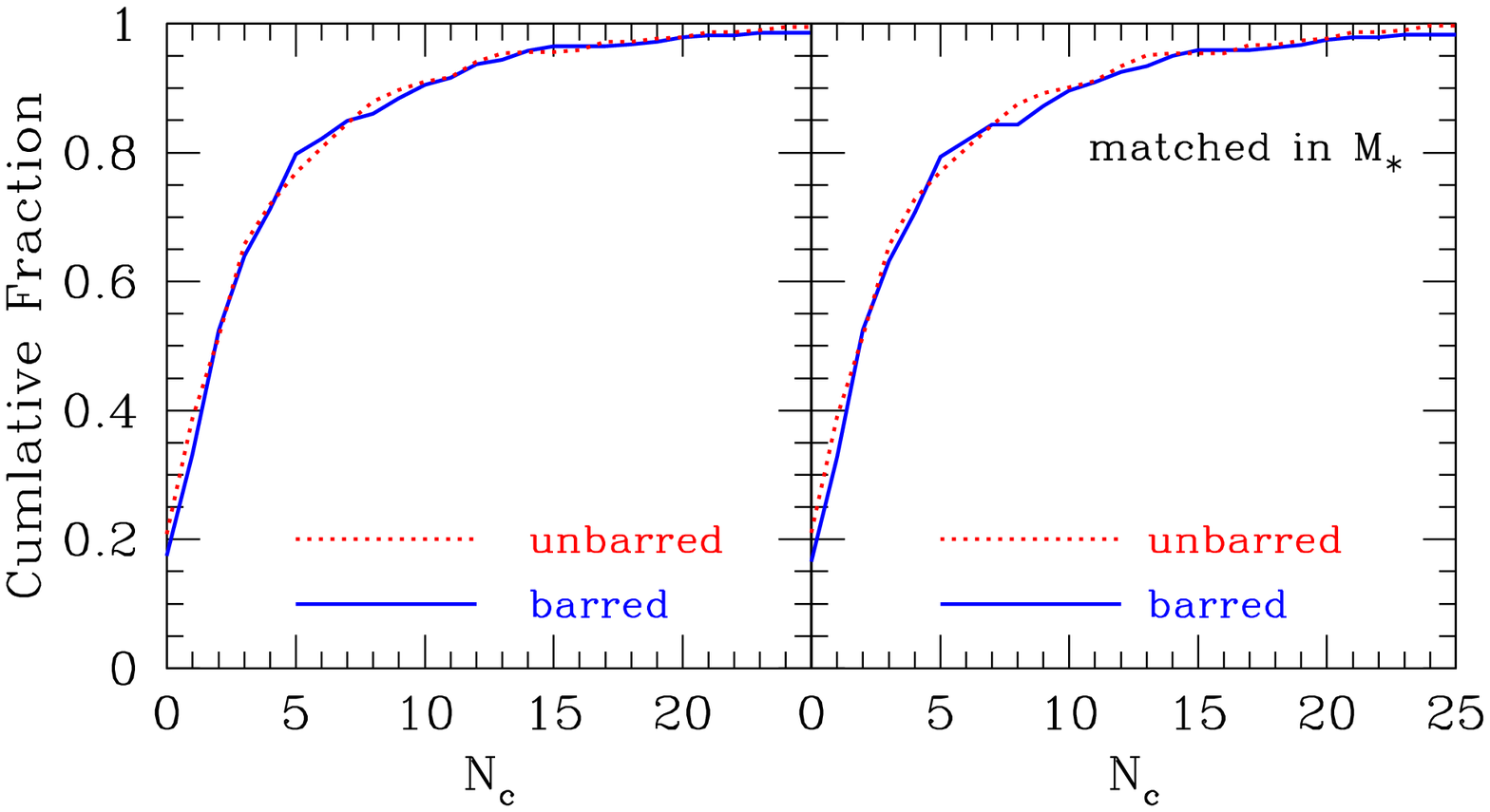,clip=true,width=0.5\textwidth}}
\centerline{\psfig{figure=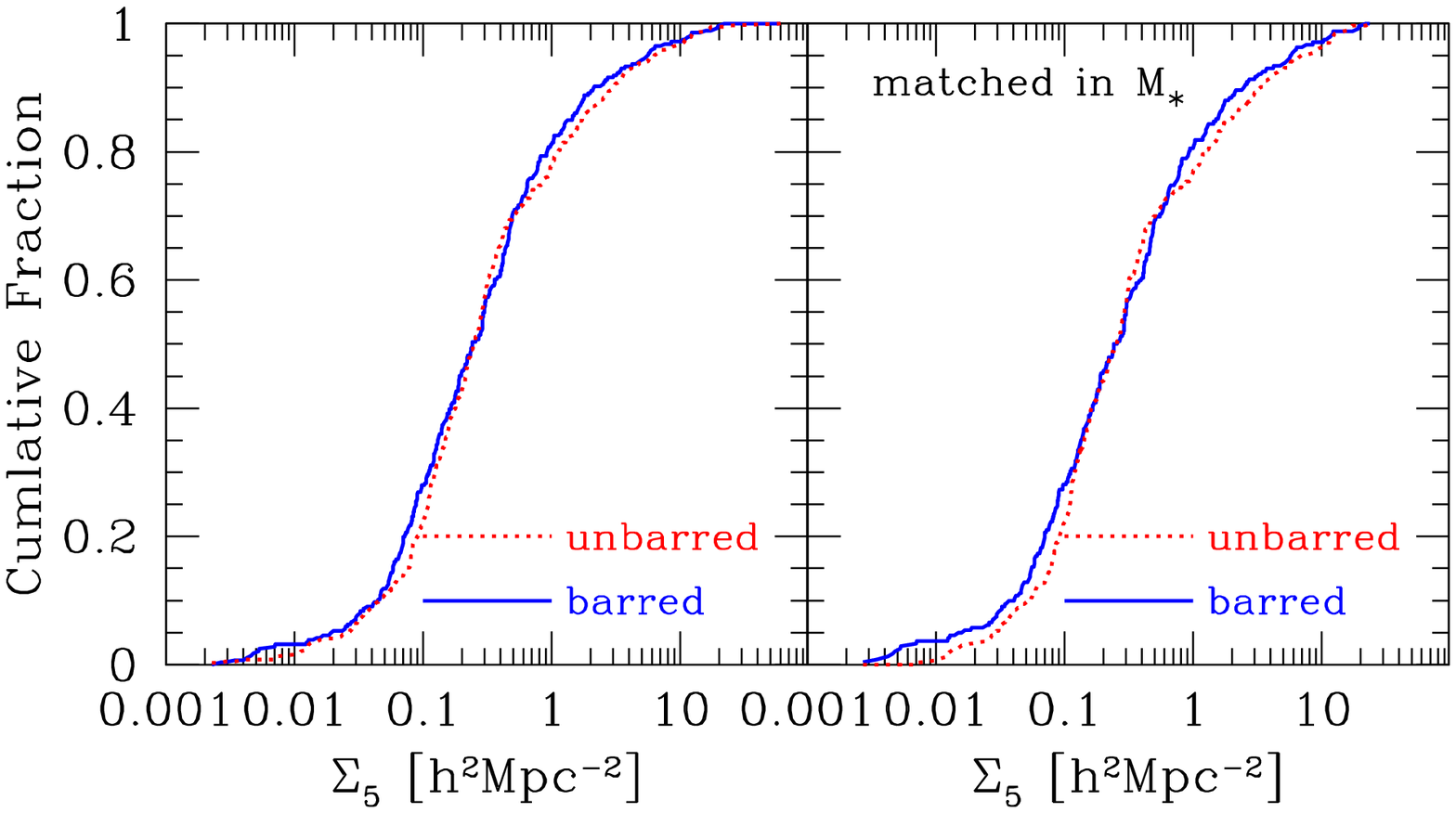,clip=true,width=0.5\textwidth}}
\caption{Cumulative  distribution  of  barred  (solid blue  line)  and
  unbarred  (dashed red  line) spiral  galaxies as  a function  of the
  number  of  neighbour galaxies  within  2  $h^{-1}Mpc$ in  projected
  radius (top panels), and the local galaxy density $\Sigma_5$ (bottom
  panels) estimated using the  projected distance to the fifth nearest
  neighbour galaxy  (see text for  details).  In the  right-hand panel
  the two samples are closely matched in stellar mass.}
\label{fig:sigma5_dist}
\end{figure}

\begin{figure}
\centerline{\psfig{figure=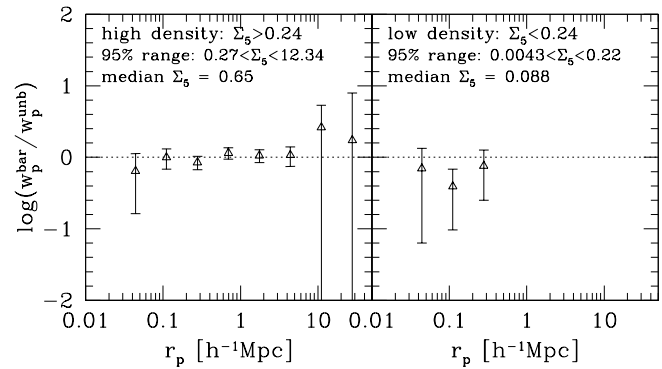,clip=true,width=0.5\textwidth}}
\caption{Ratio  of $w_p(r_p)$  of bars  relative to  that  of unbarred
  spirals.  The left (right) panel  is for galaxies with local density
  $\Sigma_5$  higher  (lower) than  0.24  $h^2$Mpc$^{-2}$, the  median
  $\Sigma_5$  of  the whole  sample.  In  each  panel the  barred  and
  unbarred samples are closely matched in stellar mass. The median and
  the 95\% range of the $\Sigma_5$ value of the bars are indicated.}
\label{fig:wrp_sigma5}
\end{figure}

The  power  of  the  $w_p(r_p)$  statistic  is  that  it  encapsulates
information about  how galaxy properties depend on  environment over a
wide  range   of  physical  scales,  including  both   the  regime  of
galaxy-galaxy interactions at  scales below 100 kpc and  the regime of
local environment at larger scales.   There have also been a number of
studies directly examining the  correlations between the fractions and
properties   of  bars   in  galaxies   and  their   local  environment
\citep[e.g.][]{Aguerri-Mendez-Abreu-Corsini-09}.        The      local
environment is  usually expressed in terms of  overdensities in galaxy
number that are  estimated in a fixed sphere/aperture  centred on the
galaxies being studied. The radius of the sphere/aperture is chosen to
be less  than a few  Mpc so as to  probe the physical  processes occurring
inside dark matter haloes.

In order to make comparisons with such studies as well as to understand
the {\em real} environments of  our galaxies, we have calculated local
density using two different quantities that have been commonly adopted
in previous  studies. We first follow  \cite{Kauffmann-04} and compute
the  number  of  galaxies  $N_c$  in the  reference  sample  within  2
$h^{-1}Mpc$ in projected radius and  $\pm 500$ km s$^{-1}$ in velocity
difference   from   the  galaxy   being   studied.   We  also   follow
\citealt{Balogh-04}                      (see                     also
\citealt{Aguerri-Mendez-Abreu-Corsini-09})  and  estimate a  projected
local  density $\Sigma_5$  of  each  galaxy in  our  sample using  the
projected distance $d_5$ to  its fifth nearest neighbour galaxy within
$\pm1000$  km   s$^{-1}$.   For  both  quantities   we  only  consider
neighbouring  galaxies  that are  brighter  than  $M_{r}=-20$ and  not
closer  than 50 $h^{-1}$  kpc.  The  bright limit  gives us  a uniform
density estimate  that is  applicable to our  magnitude-limited sample
over the  full redshift  range probed. The  lower limit  for projected
distance is chosen such that the  galaxy itself is not included in the
density estimate  and, more importantly,  the density estimate  is not
biased by  SDSS fibre collisions.  It  is important to  note that, for
these density calculations, we have constructed a new reference sample
using the  SDSS data  release 7 \citep[DR7;][]{SDSS-DR7}  which covers
much larger  area on  the sky than  do our samples  of barred/unbarred
galaxies. By  this way we  have avoided for  most of our  galaxies the
potential bias  in the estimated densities  due to edges  and holes in
the survey.

Figure~\ref{fig:sigma5_dist} shows  the cumulative fraction  of barred
and  unbarred galaxies in  our sample  as a  function of  $N_c$ (upper
panels)  and $\Sigma_5$  (lower  panels).  Results  are  shown in  the
left-hand  panels for  the  original samples,  and  in the  right-hand
panels   after  the  two   samples  have   been  matched   in  stellar
mass. Overall, only 20 per cent of our galaxies have no neighbours and
around half  have two or more  neighbours.  These numbers  are in good
agreement with those quoted in \citet{Kauffmann-04} (see their Fig.~1).
Our   estimates  of   $\Sigma_5$  also   agree  well   with   that  of
\citet{Aguerri-Mendez-Abreu-Corsini-09} (see their Fig.~10).  Around 80
per cent of the galaxies are  located in underdense regions with $\Sigma_5<1$
$h^{2}$Mpc$^{-2}$.   The median value  of $\Sigma_5$  is 0.24  for the
galaxies as a whole.

Figure~\ref{fig:sigma5_dist}  shows  that   there  is  no  significant
correlation between  the presence of a  bar in the  galaxies and their
local densities. This is true no matter how the density is quantified
and whether or  not the two comparison samples  are matched in stellar
mass.     Our    results    again    are    well    consistent    with
\citet{Aguerri-Mendez-Abreu-Corsini-09},  although  they have  treated
the edge  effect in a more  careful way.  After  excluding galaxies at
less than 7  Mpc from the nearest edge of SDSS,  the authors still did
not find any difference between the environment of barred and unbarred
galaxies.

In   Figure~\ref{fig:wrp_sigma5}  we   compare  the   projected  2PCCF
$w_p(r_p)$ for  barred and unbarred  samples, matched in  stellar mass
but  with  different local  densities.  For  this  we divide  all  the
galaxies into two subsamples according  to whether they are located in
regions with $\Sigma_5$ higher or lower than 0.24 $h^2$Mpc$^{-2}$, the
median value of the whole  sample. For each subsample, we then compute
$w_p(r_p)$ for the barred  and unbarred galaxies separately.  Again we
have trimmed the barred and unbarred samples so that they show similar
stellar mass  distributions. As can be  seen from the  figure there is
still  no difference  between the  clustering of  barred  and unbarred
galaxies on all scales where we have $w_p(r_p)$ measurements.

In  the last  two  columns in  Figure~\ref{fig:wrp_bars}  we show  the
results for bars  with different properties, but for  those located in
high and low densities separately. There is no significant effect anywhere.

\section{Summary and Discussion}
\label{sec:summary}

We  have  used  a  sample  of  nearly  1000  galaxies  (including  255
ellipticals,  389 unbarred  and  286 barred  spirals)  from the  Sloan
Digital Sky  Survey (SDSS) with  well-determined structural parameters
to study the clustering properties  of galaxies with bars  in the  
low-redshift Universe.  For the first time this  is quantified  by the
projected  redshift-space two-point  correlation  function $w_p(r_p)$,
which is computed over a range of scales from $\sim 10$ kpc to a $\sim
10$ Mpc. In  practice, we cross-correlate the sample  of galaxies with
well-determined structural parameters with a large reference sample of
galaxies drawn  from full  DR4 spectroscopic sample.   The statistical
errors on  our results  are acceptable  due to the  large size  of the
reference sample.

The measurement  of $w_p(r_p)$  on all scales  does not depend  on the
presence or absence of a bar.   This is true when the bars are divided
into subsamples according to  their optical colours, a crude indicator
of mean stellar  age and possibly bar dynamical age, or when the host  
galaxies are divided according
to  stellar mass  or  bulge-to-total luminosity  ratio.  We have  also
examined the dependence of clustering on the properties of bars. These
include the bar-to-total luminosity  ratio, bar morphology measured by
ellipticity   and  boxyness,  and   the  surface   brightness  profile
characterised by  the effective surface brightness,  length and Sersic
index.  In addition, we  checked the  clustering dependence  on global
galaxy properties such  as size and mean stellar  surface density.  We
find a stronger hint that the clustering depends  only on ellipticity.  
Bars with high ellipticities appear to be more strongly clustered than  
those  with  low ellipticities on scales between $\sim$  1 and 10 
$h^{-1}$Mpc, and this is true only for red bars.

On scales larger than a  few Mpc, the amplitude of $w_p(r_p)$ directly
measures the  mass of  the dark matter  haloes that host  the galaxies
\citep[e.g.][]{Jing-Mo-Boerner-98}.  Our results thus suggest that the
formation and  evolution of bars are  independent of the  mass of dark
matter haloes in  which their host galaxies are  found.  The amplitude
of $w_p(r_p)$ on  scales $\la 100$ kpc probes  physical processes such
as mergers  and interactions \citep[e.g.][]{Li-08a,Li-08b}.   Thus our
results also  suggest that the formation  of bars is  not dominated by
external processes such as interactions with close companions.

Our results  provide support  that bars in  the local Universe  may be
predominantly    produced   by   global    gravitational   instability
(\citealt{Toomre-1981,  Sellwood-93}). Numerical  simulations indicate
that  bars  can  form  spontaneously  in galactic  discs,  usually  on
relatively   short    timescales.   There   is  a   hint  of
environmental effects  on bar properties, for  example the ellipticity
of red  bars which may be  a consequence of the  fact that bar ellipticity
tends to increase with the age of the galaxy.

We now list a number of  important caveats. First, our sample of bars,
while larger than most previous studies, is still relatively small and
the errors on our correlation function measurements are larger than we
would like.  Our  sample is also biased towards  strong (thus possibly
old) bars  and high-luminosity galaxies,  and so our  conclusions must
also be limited to these galaxies.  The blue bars in our sample, which
are        presumably       ``young''       \citep{Gadotti-deSouza-05,
  Gadotti-deSouza-06},  may be already  old enough  and we  might have
missed  the  connection  with  close  companions  due  to  this  bias.
Finally, galaxies associated with  ongoing major mergers were excluded
from our (visually selected) sample and so our $w_p(r_p)$ measurements
are probably biased low at  very small separations. The effect of this
can be seen from the small scale correlation functions.  In all of our
figures, we don't have $w_p(r_p)$ measurements at $r_p$ below $\sim$40
kpc, indicating  that we  have been only  able to  probe galaxy-galaxy
interactions  with projected  separations larger  than this  scale and
have lost all closer systems.

Our results  are consistent with  \citet{vdB-02} who used  the Palomar
Sky  Survey to  classify 930  galaxies into  field, group  and cluster
environments,  and  found that  $25\%\pm  3\%$,  $19\%  \pm 4\%$,  and
$28\%\pm 3\%$ of  galaxies were barred (i.e. there  was no significant
trend  with  environment).   Our  results  are  also  consistent  with
\citet{Aguerri-Mendez-Abreu-Corsini-09} who  examined the relationship
between  bar  properties and  local  galaxy  density  using 2106  disc
galaxies from the SDSS and  also concluded that the properties of bars
do  not  depend on  the  local  environment.  \citet{Barazza-09}  have
recently compared the properties of bars in field and clusters using a
sample  of 925 galaxies  at redshifts  $z=0.4$--$0.8$, and  found that
bars   in  clusters   are   slightly  longer   than   that  in   field
galaxies.  However, given  the  different redshift  ranges, the  small
sample sizes,  and the  different sample selections,  this discrepancy
should not be  overemphasised and needs to be  revisited in the future
with larger samples.

With the large number of galaxies in the SDSS, a much larger sample of
barred galaxies can be  constructed using more objective criteria (see
e.g. \citealt{Barazza-Jogee-Marinova-08}),  including fainter galaxies
and  weaker bars. When  such a  sample becomes  available, it  will be
possible  to have  better  statistics and  more  reliable results.  We
intend to return to this topic in our  future work.

%%%%%%%%%%%%%%%%%%%%%%%%%%%%%%%%%%%%%%%%%%%%%%%%%%%%%%%%%%%%%%%%%%%%%%
\section*{Acknowledgments}
We thank Lia Athanassoula for  useful discussions and the referee for
helpful comments.  CL is supported by
the  Joint Postdoctoral  Programme in  Astrophysical Cosmology  of Max
Planck   Institute   for   Astrophysics  and   Shanghai   Astronomical
Observatory,   by   NSFC   (10533030,   10633020),  by   973   Program
(No.2007CB815402)  and  by the  Knowledge  Innovation  Program of  CAS
(No.KJCX2-YW-T05).    DAG    is     supported    by    the    Deutsche
Forschungsgemeinschaft  priority program  1177 (``Witnesses  of Cosmic
History: Formation  and evolution of  galaxies, black holes  and their
environment''), and  the Max Planck  Society.  SM thanks  the Humboldt
Foundation  for  travel  support  and  the  Max-Planck  Institute  for
Astrophysics for hospitality.

Funding  for the  SDSS and  SDSS-II has  been provided  by  the Alfred
P.  Sloan  Foundation, the  Participating  Institutions, the  National
Science  Foundation,  the  U.S.  Department of  Energy,  the  National
Aeronautics and Space Administration, the Japanese Monbukagakusho, the
Max  Planck Society,  and  the Higher  Education  Funding Council  for
England. The SDSS Web Site is http://www.sdss.org/.

The SDSS is  managed by the Astrophysical Research  Consortium for the
Participating  Institutions. The  Participating  Institutions are  the
American Museum  of Natural History,  Astrophysical Institute Potsdam,
University  of Basel,  University of  Cambridge, Case  Western Reserve
University,  University of Chicago,  Drexel University,  Fermilab, the
Institute  for Advanced  Study, the  Japan Participation  Group, Johns
Hopkins University, the Joint  Institute for Nuclear Astrophysics, the
Kavli Institute  for Particle  Astrophysics and Cosmology,  the Korean
Scientist Group, the Chinese  Academy of Sciences (LAMOST), Los Alamos
National  Laboratory, the  Max-Planck-Institute for  Astronomy (MPIA),
the  Max-Planck-Institute  for Astrophysics  (MPA),  New Mexico  State
University,   Ohio  State   University,   University  of   Pittsburgh,
University  of  Portsmouth, Princeton  University,  the United  States
Naval Observatory, and the University of Washington.

%%%%%%%%%%%%%%%%%%%%%%%%%%%%%%%%%%%%%%%%%%%%%%%%%%%%%%%%%%%%%%%%%%%%%%
\bibliography{barredsample,bars,bars2,sdss,mypub}

\bsp
\label{lastpage}
\end{document}

%% file: macro1.tex
\newcommand{\beq}{\begin{equation}}
\newcommand{\eeq}{\end{equation}}

\newdimen\hssize
\newdimen\hdsize 